\documentclass[reprint,amsmath,amssymb,showkeys,aps,prb]{revtex4-1}
\usepackage{natbib}
\usepackage{graphicx}
\usepackage{dcolumn}
\usepackage{bm}
\usepackage{hyperref}
\usepackage{hhline}
\begin{document}
\title{Finite-temperature phase diagram and critical point of the Aubry pinned-sliding transition in a $2D$ monolayer}
\author{
        Davide Mandelli$^{1,2}$, 
        Andrea Vanossi$^{3,1}$, 
        Nicola Manini$^{4,1,3}$, 
        Erio Tosatti$^{1,3,5}$
       }

\affiliation{
$^1$ International School for Advanced Studies (SISSA),
     Via Bonomea 265, 34136 Trieste, Italy \\
$^2$ Tel Aviv University, Tel Aviv 6997801, Israel,\\
$^3$ CNR-IOM Democritos National Simulation Center,
     Via Bonomea 265, 34136 Trieste, Italy \\
$^4$ Dipartimento di Fisica, Universit\`a degli Studi di Milano, 
     Via Celoria 16, 20133 Milano, Italy \\
$^5$ International Centre for Theoretical Physics (ICTP),
     Strada Costiera 11, 34014 Trieste, Italy
}
\date{\today}
\begin{abstract}
The Aubry unpinned--pinned transition in the sliding of two
incommensurate lattices occurs for increasing mutual interaction
strength in one dimension ($1D$) and is of second order at $T=0$,
turning into a crossover at nonzero temperatures.
Yet, real incommensurate lattices come into contact in two dimensions
($2D$), at finite temperature, generally developing a mutual
Novaco-McTague misalignment, conditions in which the existence of a
sharp transition is not clear.
Using a model inspired by colloid monolayers in an optical lattice as a
test $2D$ case, simulations show a sharp Aubry transition between an
unpinned and a pinned phase as a function of corrugation.
Unlike $1D$, the $2D$ transition is now of first order, and,
importantly, remains well defined at $T>0$.
It is heavily structural, with a local rotation of moir\'e pattern
domains from the nonzero initial Novaco-McTague equilibrium angle to
nearly zero.
In the temperature ($T$) -- corrugation strength ($W_0$) plane, the
thermodynamical coexistence line between the unpinned and the pinned
phases is strongly oblique, showing that the former has the largest
entropy.
This first-order Aubry line terminates with a novel critical 
point $T=T_c$, marked by a susceptibility peak.
The expected static sliding friction upswing between the unpinned and
the pinned phase decreases and disappears upon heating from $T=0$ to
$T=T_c$.
The experimental pursuit of this novel scenario is proposed.
\end{abstract}
\pacs{pacs}
\maketitle
\section{Introduction}
\label{sec:Intro}
There is in tribology -- the science of friction and adhesion -- a long standing interest in models of
periodic solid interfaces that are incommensurate, where the crystal unit
cells facing each other are fundamentally mismatched in either size or
angle or both, so that they cannot fit in a common unit cell of any finite
size.
While in the macroscopic world it is generally difficult to
realize a perfectly incommensurate interface because of defects,
irregularities and temperature, in nanotribology\cite{vanossiRevModPhys,Manini16} 
there are well-defined nanoscale
realizations of incommensurate crystal interfaces, such as graphene and
other $2D$ sheets,\cite{dienwiebel2004,leven2013} rare-gas
monolayers,\cite{pierno2015} and colloid monolayers in an optical
lattice.\cite{bohlein2012,vanossi2012}
The present work is devoted to understand the relative state of pinning of
an idealized, yet well defined realization of this type of $2D$ contact,
focusing especially on its evolution under conditions of finite
temperature, a question totally unexplored so far.

We begin with a brief review, which starts from the $1D$ Frenkel-Kontorova
(FK) model,\cite{Floria96,braun_kivshar} consisting of a harmonic chain of
classical point particles in a static sinusoidal potential of amplitude
$W_0$, which acts as a corrugation opposing chain sliding.
Incommensurability between the mean interparticle spacing $a_c$ and the
sinusoidal potential wavelength $a_p$ occurs when the ratio $\rho=a_c/a_p$
is irrational.
With respect to the unperturbed chain, the incommensurate potential causes
a distortion of the particle positions that can be described by a
deformation of the chain's local phase $\Phi(x)$ relative to the reference
phase of the corrugation sinusoid.
As the chain-sinusoid interaction increases from the non-interacting
straight behavior $\Phi_0(x)=(\rho-1)x$ at $W_0=0$, the phase deforms into
a smooth staircase shape with the same mean slope $(\rho-1)$, but now
sporting nearly commensurate and horizontal steps where $\Phi(x)$ is
approximately constant, separated by jumps, called solitons or misfit
dislocations, or kinks, and antisolitons (antikinks), where most of the
misfit stress associated with $\nabla \Phi$ is concentrated.
Incommensurability implies that the total energy $E$ of the chain is in all
cases strictly independent of its center-of-mass position,
$\delta E /\delta \Phi=0$.
Because of this, it had long been held that the translational dynamics of
an incommensurate interface should always be gapless -- causing in this
case the chain state to be unpinned and thus shiftable by an arbitrarily
small force.
Back in the 1980s Aubry proved mathematically that this is not always so.
The $1D$ incommensurate FK model displays at $T=0$ 
and for fixed harmonic spring constant a sharp phase
transition\cite{aubry1983} between an unpinned phase where the corrugation
is weak (soliton widths larger then or comparable to that of the steps,
  small overall distortion $\Phi(x)-(\rho-1)x$), and a pinned phase, realized above a
critical corrugation magnitude $W_{0c}$, where the distortion is large,
solitons are narrow and the phase distortion is large.
Above $W_{0c}$ -- whose magnitude depends on precise parameters and on
incommensurability -- the chain develops a nonzero gap against sliding.

The well-known physical essence of the Aubry transition is that the
relative probability to find a particle exactly at a potential maximum --
probability which is finite so long as the chain is unpinned -- drops
mathematically to zero at $W_0>W_{0c}$, constituting a self-generated constraint
to the chain center-of-mass dynamics and to its sliding motion at
$T=0$.\cite{coppersmith1983}
That is, despite
$\delta E /\delta \Phi=0$, the dynamical constraint limits
phase-space accessibility, effectively breaking ergodicity and causing the
onset of static friction -- pinning of the chain against free sliding under
an infinitesimally small external force.
Even if it is, for any finite system size and within mean-field, a regular
structural and thermodynamic transition,\cite{mazzucchelli1985} the Aubry
transition does not possess a proper Landau-type order parameter.
It has instructively been characterized by a ``disorder parameter''
measuring the extension of the forbidden phase space, a concept of which we
will make use later.
Even if the step-terrace deformation of the chain's phase $\Phi(x)$ retains
a qualitatively similar nature to that in the unpinned state, 
the solitons evolve from broad and overlapping to non-overlapping and atomically sharp, and
both static and dynamic friction change at the transition. In the unpinned phase, the
chain slides under any applied force however weak, leading to a state of
flow of the solitons.
This absence of static friction is sometimes referred to as
``superlubricity'' or ``structural lubricity'' in the friction community.
In the pinned state, chain sliding requires a static friction threshold
force to be overcome, before sliding sets in.

The $D=1$, $T=0$ physics being well understood, it does remain rather
academic unless it can be brought closer to the real, in our case
nanotribological, world.\cite{benassi2011,mandelli2013,gangloff2015,bylinskii2015} 
Real incommensurate lattices come into contact at
an interface which is $D=2$ dimensional.
Moreover, temperature is always finite and often large.
Although the physics of the unpinned-pinned transition\cite{braun_kivshar}
is usually and reasonably assumed to be the same in $D=2$ and $T>0$ as that
of Aubry's case in $D=1$ and $T=0$ that assumption is neither theoretically
proven nor at least demonstrated in a specific experimentally relevant
example.
Clearly, the two regimes of essential unpinning and of strong pinning must
surely exist for flat $2D$ incommensurate contacts.
But whether there is or not between them a sharp transition as a function
of the contact strength, and if so precisely what kind of transition, must
still be determined.
At $T=0$, moreover, the $1D$ Aubry transition is continuous and critical as
a function of mechanical parameters.
Although in mean-field theory the second-order Aubry transition may extend
to finite temperature,\cite{mazzucchelli1985} in reality in $1D$ the sharp
phase transition is strictly limited to $T=0$, turning into a
smooth crossover at any finite temperature.\cite{braun_kivshar}

A recent study\cite{mandelli2015prb} of a $2D$ model colloid monolayer in
an incommensurate optical lattice provides the first interesting example of
such a transition in a real class of systems where experiments are actively
going on.\cite{bohlein2012,Reichhardt16}
That study revealed at $T=0$ a $2D$ Aubry-type unpinned-pinned transition
which is, unlike the $1D$ case, of first order for increasing corrugation.
In that transition the Novaco-McTague misalignment angle,\cite{novaco1977}
a specific $2D$ feature, plays an important role.
Moreover, the two components of the total (potential) energy, namely the
interparticle and the particle--substrate-potential terms, undergo opposite
and compensating jumps.
The evolution of this $2D$ unpinning-pinning transition in the real finite
temperature situation remains as yet unknown.

Here we show, exploiting the same $2D$ colloid monolayer/optical lattice
model as a relevant system which we can study by molecular
dynamics (MD) simulations, that the Aubry-like transition remains well
defined and of first order at nonzero temperature, where it gives rise to a
clear phase line between the unpinned and the pinned states.
The large positive slope of this line indicates via Clausius-Clapeyron's
equation that the unpinned phase has the largest entropy, revealing some
collapse of accessible phase space in the pinned phase.
A disorder parameter, the $2D$ version of Coppersmith-Fisher's $1D$
one,\cite{coppersmith1983} is correspondingly identified, and its jump is
demonstrated at the onset of pinning.
From the geometric viewpoint, the $2D$ Aubry transition in this model
system is heavily structural.
Its main feature is a rotation of {\it local moir\'e pattern domains} from the
nonzero Novaco-McTague-like misalignment angle in the weak-corrugation
unpinned phase to nearly zero in the strong-corrugation pinned phase.
The two phases appear to possess the same spatial symmetry. Thus the
first-order line is accordingly expected and actually found to terminate at
a high-temperature critical point, where unpinned and pinned characters are
lost and therefore merge.
Limitations of simulation size and time do not permit here a
characterization of this critical point, which qualitative considerations
would tentatively place in the universality class of the gas--liquid
transition.
As in liquid-gas, the particle-particle energy and the entropy jumps
contributing to the free energy are equal in magnitude at the transition:
however in this Aubry case they have the {\em same sign} rather than
opposite sign, their positive sum exactly compensating the gain of periodic
potential energy (a term absent in liquid-gas).

After this characterization of equilibrium properties, we address
tribological and dynamical questions by carrying out
further simulations and extracting static friction under an
external force.
The change from zero to finite static friction characterizes the unpinned
and pinned nature respectively, confirming a change from lubricity of the
unpinned phase to sticking of the pinned phase which persists at $T>0$,
and only vanishes at the terminal critical temperature.
\section{Model and simulations}
\label{sec:Mod&Prot}
Following our previous $T=0$ work\cite{vanossi2012,mandelli2015prb} we
describe the colloidal particles as classical point objects interacting via
a screened repulsive Coulomb potential
\begin{equation}
\label{eq:Yuk}
 V(r)=\frac{Q}{r}\exp\left(-r/\lambda_{\rm D}\right),
\end{equation}
where $r$ is the interparticle distance, $Q$ is the coupling strength, and
$\lambda_{\rm D}$ is a Debye screening length.
Particle motion is restricted to two dimensions, where unperturbed colloids
form a triangular lattice of spacing $a_c$ in the $(x,y)$ plane. The
externally added $2D$ periodic triangular corrugation potential
\begin{eqnarray}
\label{eq:tripot}
W({\bf r})&=&-W_0\frac{2}{9}\left[\frac{3}{2}+2\cos\frac{2\pi x}{a_l}\cos\frac{2\pi y}{\sqrt{3}a_l}
            +\cos\frac{4\pi y}{\sqrt{3}a_l}\right]\nonumber\\
          &=&W_0w({\bf r})
\end{eqnarray}
has strength $W_0$ and periodicity $a_l$, representing the experimental
optical lattice.
Here we restrict particle motion to $2D$, even if
real colloids can move in $3D$,
because we are interested in the $2D$ problem in the first place. Moreover,
the experimental setup\cite{bohlein2012} implements an additional strong
confining laser force in the $z$ direction, which suppresses drastically all
vertical fluctuations to less than 5\% of the particle diameter.
The total potential energy of $N_p$ particles is
\begin{equation}
\label{eq:Hpbc}
H= \sum_{i=1}^{N_p}  \left[ W({\bf r}_i) + \frac 12 \sum_i\sum_{j \neq i} V(r_{ij}) \right].
\end{equation}
The $j$-th particle displacement ${\bf r}_j$ obeys the Langevin equation
\begin{equation}
\label{eq:dyn}
 m\ddot{{\bf r}}_j+\eta\dot{{\bf r}}_j=
-\nabla_{{\bf r}_j}\left[ \sum_{i\neq j} V(r_{ij})+W({\bf r}_j) \right]+\xi R(t),
\end{equation}
where $m$ is the mass, $\eta$ is the viscosity of the solvent, and $R(t)$ is a delta-correlated stationary Gaussian-process, satisfying
\begin{equation}
\langle R(t)\rangle=0
\end{equation}
\begin{equation}
\langle R(t)R(t')\rangle=\delta(t-t').
\end{equation}
The fluctuation-dissipation theorem is satisfied by setting $\xi=\sqrt{2\eta k_B T}$, where $k_B$ is Boltzmann's 
constant and $T$ is the temperature.
An overdamped dynamics of these particles is generated by integrating the
equations of motion with a large viscous coefficient $\eta=28$ and
$Q=10^{13}$, $\lambda_{\rm D}=0.03$.
In our simulations we further assume $2D$ periodic boundary conditions (PBCs).
Results are expressed in terms of the same system of units defined
in Table~I of Ref.~\onlinecite{mandelli2015prb}.
In these units, very roughly inspired by experimental
systems,\cite{bohlein2012} $T\sim0.04$ corresponds to room temperature.

Incommensurability between the particle monolayer and the $2D$ periodic
potential generally arises both from their different lattice spacing and
from their relative misalignment (rotation) angle.
Any chosen misalignment angle between the corrugation $W({\bf r})$ and the colloidal
lattice can be implemented by means of a suitably chosen supercell, as
follows.\cite{trambly2010}
The two lattices are defined by the pairs of primitive vectors 
$\mathbf{a}_1=a_{\rm l}(1,0)$, $\mathbf{a}_2=a_{\rm l}(0.5,\sqrt{3})$, and 
$\mathbf{b}_1=a_{\rm c}(\cos\theta,\sin\theta)$,
$\mathbf{b}_2=a_{\rm c}(\cos(\theta+\pi/3),\sin(\theta+\pi/3))$.
An arbitrary supercell-periodic structure, meant to approximate the real
incommensurate system, is realized when four integers are found that
satisfy the matching condition
$n_1\mathbf{a}_1+n_2\mathbf{a}_2=m_1\mathbf{b}_1+m_2\mathbf{b}_2$.
The supercell is a larger triangular lattice of size
$L=|m_1\mathbf{b}_1+m_2\mathbf{b}_2|$, containing a total number of
particles $N_p=m_1^2+m_1m_2+m_2^2$.
We fix $a_{\rm c}=1$ and vary $n_{1,2}$, $m_{1,2}$ in search of structures
with a mismatch $\rho=a_{\rm l}/a_c\approx3/(1+\sqrt{5})\simeq0.927$ -- close
to the experimental values of Ref.\onlinecite{bohlein2012} -- and $\theta$
near the desired value, with the obvious additional constraint that the
number of particles $N_p$ should not be too large.
We consider in practice the aligned configuration $\theta=0$, plus
misaligned configurations: $\theta\simeq 5^\circ$, $\theta\simeq 10^\circ$
and $\theta_{\rm opt}\simeq2.54^\circ$.
The latter is close to the (Novaco-McTague) equilibrium misalignment angle
$\theta_{\rm NM}\simeq2.58^\circ$ predicted by weak-coupling elastic theory
for the present parameters and
$\rho=0.927$.\cite{novaco1977,mandelli2015prl}
We could have equally chosen to study an overdense case, $\rho>1$. However, the
underdense regime $\rho<1$ which we have chosen is better suited because
overdense local compressions may favor large bucklings of particles out of
the plane, and because the energetics (not symmetrical with respect to
$\rho>1$) is less convenient in that case.
The supercell parameters adopted here are the same as in
Ref.~\onlinecite{mandelli2015prb}.
\begin{figure}[!t]
 \begin{center}
 \includegraphics[angle=0, width=0.45\textwidth]{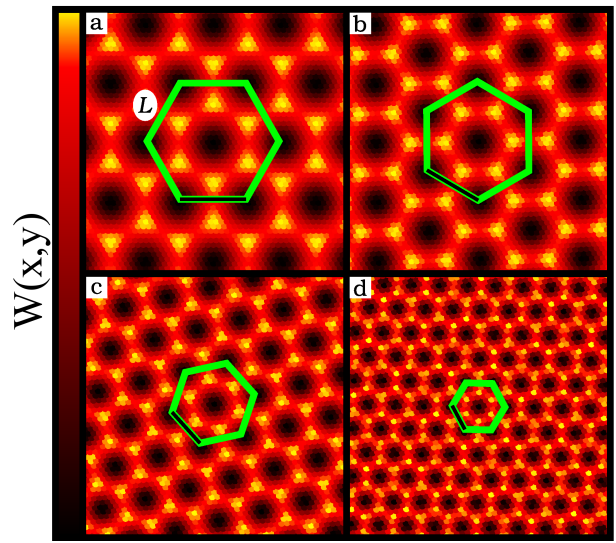}
 \caption{\label{fig:moire-examples}(Color online)
   Examples of the moir\'e patterns obtained at $\rho\simeq0.927$ for
   misfit angles (a) $\theta=0$, (b) $\theta_{\rm
     opt}\simeq2.54^\circ$, (c) $\theta\simeq5^\circ$, and (d)
   $\theta\simeq10^\circ$.
   Each dot represents a particle, whose color reflects the local 
   corrugation potential $W(x,y)$: dark for potential minima, bright for
   maxima.
   A small portion of the simulation supercells is displayed, containing an
   {\em undistorted} monolayer, at $W_0=0$.
   According to Eq.~\eqref{eq_alpha}, at $\theta_{\rm opt}\simeq 2.54^\circ$,
   panel (b), the moir\'e orientation is $\alpha\simeq 30^\circ$.
   As $\theta$ increases beyond $\theta_{\rm opt}$, the superstructure
   periodicity $L$ shrinks rapidly and rotates all the way to $\alpha \simeq
   60^\circ$ (d).
 }
 \end{center}
\end{figure}

\begin{figure}[!t]
 \begin{center}
 \includegraphics[angle=0, width=0.45\textwidth]{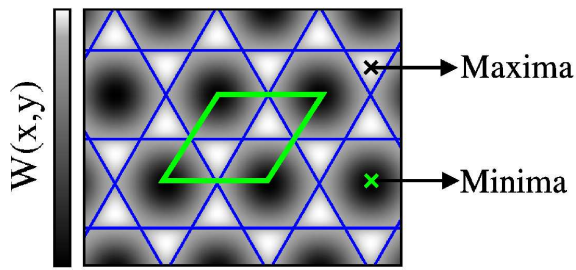}
 \caption{(Color online)
   Color map of the $2D$ periodic triangular substrate potential
   $W(x,y)$. A primitive cell is highlighted at the center.
   Isolines are drawn corresponding to the saddle-point value separating
   adiacent minima.
 }
 \label{fig:papillon}
 \end{center}
\end{figure}

The mismatch between the $2D$ particle lattice and the periodic potential
produces a moir\'e pattern corresponding to a superlattice of hexagonal
domains where particles and potential are mutually nearly commensurate,
separated by a honeycomb network of (anti-)soliton lines whose thickness
decreases with increasing corrugation strength.
Examples of the moir\'e superstructures are shown in
Fig.~\ref{fig:moire-examples}.
We recall here for clarity the relation\cite{bohr1992} between the
misalignment angle $\theta$ and the moir\'e orientation $\alpha$
\begin{equation}
\label{eq_alpha}
\cos{\theta}=\rho^{-1}\sin^2{\alpha}+\cos{\alpha}\sqrt{1-\rho^{-2} \sin^2{\alpha}} \,.
\end{equation}
The moir\'e pattern visually underlines the difference between particles
whose position is near the energetically favorable potential minima, and
others near the unfavorable potential maxima.
In the $1D$ case studied in the 1980s, the presence or absence of pinning
was described by a ``disorder parameter'' $\Psi$, roughly measuring the
radius of the neighborhood of each potential maximum which, at $T=0$,
turned from (partly) occupied in the unpinned state to exactly empty in the
pinned state.\cite{coppersmith1983}
The (tribological) essence of the Aubry transition is that when all states are accessible
and the disorder parameter is zero, the incommensurate system is unpinned
and can slide under an arbitrarily small force, whereas when the occupancy
of potential maxima and their neighborhood drops to zero and therefore the
disorder parameter is nonzero, the system is pinned and free sliding is
impeded.
Looking for a $2D$ analog of the disorder parameter we measure\cite{mandelli2015prb}
the fraction $\Psi =N_s/N_p$ of particles which populate the geometrically defined, 
papillon-shaped area where the periodic potential $W(x,y)$ is repulsive, exceeding its saddle
point value (see Fig.~\ref{fig:papillon}).
The value of $\Psi(T,W_0)$ and especially its jump will be used to
characterize phase boundaries in the $(W_0,T)$ plane phase diagram.

The phase transition is studied as follows.
At each fixed and nonzero value of $T$ we carry out two series of
simulations, the first starting from an unpinned configuration at an
adequately small value of $W_0$, which is then increased in small steps
$\Delta W_0$, the second starting from a pinned configuration at an
adequately large value of $W_0$, which is then decreased in a similar
fashion.
We generally adopt $\Delta W_0=0.005-0.01$, reducing it to $\Delta W_0=0.001$
in the parameter region straddling the transition.
For each value of $W_0$ the MD simulation time is chosen so as to ensure
thermal equilibrium.
Each run at $W_0\pm\Delta W_0$ is started from a configuration equilibrated
at the previous step in the sequence.
Thermal equilibration is checked by monitoring the disorder parameter
$\Psi$, which generally increases/decreases with simulation time,
eventually reaching a plateau.
We could in principle have used the convergence of another mechanically defined variable, such as, e.g.,  the internal energy, to monitor equilibration. However, internal energy fluctuations, related to specific heat, are harder to handle than those of the disorder parameter, which turns out to be a better choice in practice.
All relevant observables are computed from time averages along the
trajectories, discarding the initial transient time. 

All equilibrium simulations described here are carried out with $2D$ PBCs,
corresponding to ``NVT'' canonical ensemble (as opposed to a NPT ensemble,
here inaccessible).
On account of the constant volume (in this case, constant area), a
first-order phase transition in general implies an intermediate two-phase
coexistence region, since the unpinned and pinned phases generally differ
in pressure as well as in disorder parameter.
The existence of two separate phase boundaries in the $(W_0,T)$ plane is
indeed signaled by two different possible stable values of the disorder
parameter $\Psi$ -- in practice by two non coincident upward and downward
jumps of $\Psi$ for increasing or decreasing potential strength $W_0$.
As it turns out for our working parameters the width $W_2-W_1$ of the
two-phase region is narrow.
Its midpoint line $W_{0}^*=(W_1+W_2)/2$
(where $W_1$ and $W_2$ are the border values)
is therefore adequately representative of an
underlying effective constant-pressure first-order phase line.
In this way we avoid the complex questions that would otherwise arise in
order to extract a constant-pressure result, a volume change being
difficult to combine with the (rigid) periodic potential and the
requirement of fixed incommensurability.

The global angular orientation of the monolayer relative to the lattice
potential is held fixed by the PBCs, independently of temperature.
While that is an assumption reflecting a computational necessity, it does
represent those experimental realizations where very large, practically
infinite colloid islands are not expected to execute global rotations.
Actually, temperature, besides smearing somewhat the
periodic potential, would nudge the overall equilibrium orientation angle in
the direction where the total-energy growth is softer.
However, calculations at $T=0$
showed\cite{mandelli2015prl,mandelli2015prb}
that the angle of minimum energy ($2.54^\circ$ in our case) is rather
independent on the potential magnitude, as also suggested by weak coupling
theory,\cite{novaco1977} making the optimal orientation angle insensitive
to thermal smearing.
Moreover, the $T=0$ total energy is rather symmetric around the minimum,
equally soft on both sides, so that no orientational nudging is expected.
It is therefore very reasonable to adopt the same optimal $T=0$ global
orientation angle independent of temperature.

The Helmholtz free-energy of the monolayer is $F=U+W-TS$, where $U=\langle
\sum_i\sum_{j \neq i}  V(r_{ij})/2\rangle$, $W=\langle \sum_i W({\bf
  r}_i)\rangle$, and $S$ is the entropy.
By crossing the first-order transition upon variable $W_0$ and constant
$T$,
\begin{equation}
\label{Delta}
\Delta U+\Delta W-T \Delta S=\Delta F=0
\,.
\end{equation}
Simulations yield directly $\Delta U$ and $\Delta W$, both of them
mechanical quantities, across the transition.
Through Eq.~\eqref{Delta} we obtain the entropy jump $\Delta S$ at the
unpinning-pinning transition.
Of course, since no thermodynamic results such as $\Delta S$ are correctly
represented at low temperatures by a totally classical simulation, it must
be understood that all results are valid only from some (small) finite
temperature upward.
In future comparisons with experimental data this will not be a problem, 
because colloidal experiments are carried out at room temperature.

Finally, the static friction force $F_{s}$ of the monolayer, which actually
determines the presence or absence of pinning, is obtained by applying a
driving force $F_d$ to each colloid, generally along a high-symmetry
direction of the laser substrate potential. Briefly, the external force is increased
in steps $\Delta F$, and for each value of the force a simulation is carried out where the duration is 
fixed in such a way that a single free particle would move by a distance of $\Delta x=5.5$ $a_{l}$. 
The monolayer is considered to be sliding (i.e. depinned) if the total displacement of its center-of-mass at 
the end of the simulation is $\Delta x_{com}>2.0$ $a_{l}$.
\begin{figure} [!t]
 \begin{center}
 \includegraphics[angle=0, width=0.45\textwidth]{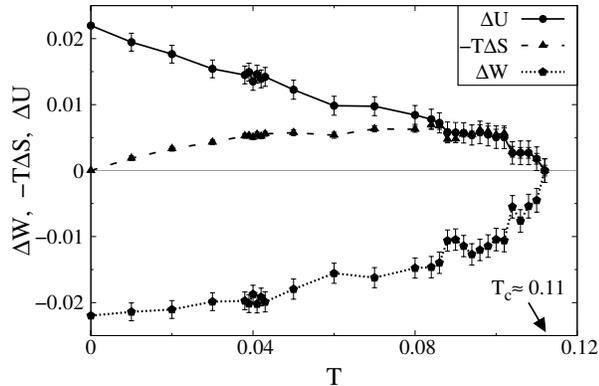}
 \caption{
   The temperature dependence of the three contributions to the free energy
   per particle evaluated across the coexistence line.
   $\Delta W$, $-T\Delta S$, $\Delta U$
   correspond respectively to the change in substrate potential energy,
   entropy, and inter-particle energy between the pinned and the unpinned
   phase.
 }
 \label{fig:deltaEnergies}
 \end{center}
\end{figure}
\section{Results: finite temperature $2D$  Aubry transition and phase diagram}
\label{sec:ResFs}
\begin{figure}[!t]
 \begin{center}
 \includegraphics[angle=0, width=0.45\textwidth]{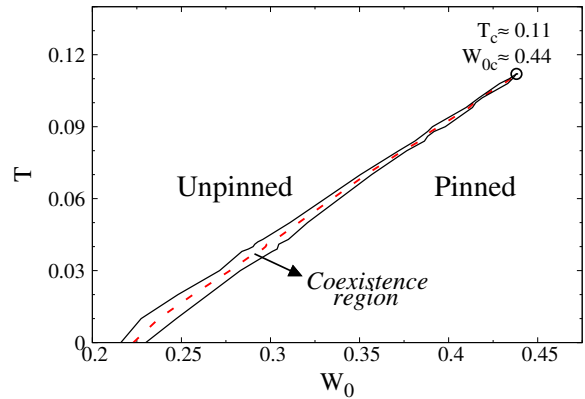}
 \caption{(Color online)
   Phase diagram in the $(W_0,T)$ plane defined by the substrate
   corrugation strength and temperature.
   At each $T$ the region of phase coexistence is confined to a very
   narrow interval ($W_1$,$W_2$) of corrugations, as shown by the
   continuous lines in the plot.
   The dashed curve shows the average value $W_0^*=(W_1+W_2)/2$, which 
   has been used to define the ``coexistence-line'' adopted in our
   thermodynamic analysis of the transition. 
   The steady slope of the coexistence line indicates that the unpinned
   phase retains consistently a higher entropy than the pinned phase.
 }
 \label{fig:phase-diagram}
 \end{center}
\end{figure}

The simulation protocol just outlined yields direct evidence that the
first-order phase transition between an unpinned phase at small corrugation
magnitude $W_0$ and a pinned phase at large $W_0$ persists at finite
temperature.
All thermodynamic quantities (except, at constant pressure, the total Gibbs free energy) jump at
the transition, as shown in Fig.\ref{fig:deltaEnergies}. Figure \ref{fig:phase-diagram} shows the phase-diagram,
 where the two-phase coexistence region (at constant volume) is very narrow,
indicating that constant volume and constant pressure are very similar. Thus Gibbs and Helmholtz
free energies only differ by a constant, and the jumps $\Delta U$, $\Delta W$ approximately coincide. 
While at $T=0$ $\Delta U=-\Delta W$ are opposite and compensate
exactly, at finite temperature entropy kicks in, and near $T_c$ the
approximate equality $-T\Delta S + \Delta U = \Delta W$ holds.
The large negative jump $\Delta W$ at pinning indicates that in the pinned
state particles benefit much more from the external potential minima.
That gain is compensated by a corresponding worsening of particle-particle 
interactions and of entropy, both much better in the unpinned phase.
Besides these reasonable outcomes, we observe in addition that at least
close to $T_c$, $-T\Delta S \sim \Delta U$, an unexpected approximate
equality for which we found no good explanation. 

The phase diagram of Fig.\ref{fig:phase-diagram} also shows that the
unpinned-pinned transition is heavily right-leaning with temperature,
ending at $W_{0c} \simeq 0.44$,  $T_c\simeq 0.11$, the latter
to be compared with $T=0.04$, the model room temperature.
The unpinned phase therefore possesses a much larger entropy than the pinned phase.
Moreover, the inverse slope $d(W_0^*/W_{0c})/d(T/T_c) = 0.50 \pm 0.05$
is quite small, in contrast with liquid-gas slopes $d(P/P_c)/d(T/T_c) \simeq
3,\ 4,\ 6.5$ for H$_2$O, a van der Waals fluid,\cite{johnston2014} and Ar,\cite{henderson1969} 
respectively.

\begin{figure}[!t]
 \begin{center}
 \includegraphics[angle=0, width=0.45\textwidth]{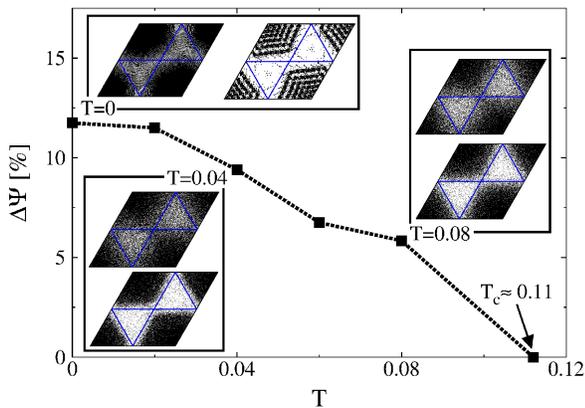}
 \caption{(Color online)
   The jump $\Delta\Psi$ of the $2D$ disorder parameter across the
   first-order phase transition is reported as a function of temperature.
   For a few values of $T$, the insets illustrate the corresponding change
   in the population of the regions above the saddle point value.
   There the positions of all particles are reported folded inside one
   primitive cell of the substrate potential (at finite $T$ several
   snapshots along the trajectory have been considered).
   Table~\ref{tab_coex} reports the two corrugation amplitudes $W_{1,2}$
   used for the insets and for the definition of $\Delta\Psi$.
   At the critical temperature $T_c\simeq0.11$ the transition becomes of
   second order and $\Psi$ varies smoothly across it: here we just set
   $\Delta\Psi(T_c)=0$ for simplicity.
 }
 \label{fig:deltaNsaddle}
 \end{center}
\end{figure}

Finally, Fig.\ref{fig:deltaNsaddle} summarizes our
resulting disorder parameter for the Aubry-type
transition of the $2D$ colloid model at $\rho = 0.927$ as a function of
temperature.
Similar to the $1D$ case,\cite{coppersmith1983} the transition is
characterized by a sudden drop of the number of particles lying near the
maxima of $W(x,y)$. This is demonstrated by the jump $\Delta\Psi$ of the $2D$ 
disorder parameter, which is finite up to $T=T_c\simeq0.11$, where it disappears.
\begin{center}
 \begin{table}[!b]
 \renewcommand{\arraystretch}{1.3}
 \begin{tabular}{  c  c  c  }
 \hhline{===}
  $T$  & $W_1$ & $W_2$ \\
  \hline
  0    & 0.216 & 0.230 \\
  0.02 & 0.247 & 0.270 \\
  0.04 & 0.288 & 0.306 \\
  0.06 & 0.331 & 0.229 \\
  0.08 & 0.371 & 0.382 \\
 \hhline{===}
 \end{tabular}
 \caption{\label{tab_coex}
   The values of the substrate potential strength $W_0$ used to define the
   jump $\Delta \Psi$ of the disorder parameter at the transition, for the
   temperatures $T$ shown in Fig.~\ref{fig:deltaNsaddle}.
   Values of $W_1$ ($W_2$) have been taken in the unpinned (pinned) phase
   immediately
   before (after) the coexistence region in the $(W_0,T)$ plane.
 }
 \end{table}
\end{center}
\section{Structural: Local Commensurate Rotation }
\label{sec:Str}
\begin{figure}[!t]
 \begin{center}
 \includegraphics[angle=0, width=0.45\textwidth]{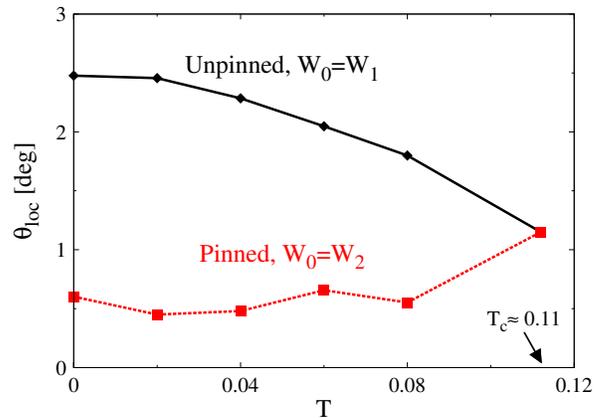}
 \caption{(Color online)
   The local angular orientation of the colloidal monolayer defined in
   Eq.~\eqref{eq:thetaloc} as a function of temperature.
   The black continuous curve shows the orientation measured at
   $W_0=W_1$ in the unpinned phase, while the red dashed curve is the
   corresponding local alignment in the pinned phase at $W_0=W_2$.
   Temperatures $T$ and corrugations $W_{1,2}$ considered here are the
   same as those of Fig.~\ref{fig:deltaNsaddle}, also reported in
   Table~\ref{tab_coex}.
   At $T=T_c\simeq0.11$ there is no discrete jump in the local
   orientation. The point reported in the plot corresponds to the average local orientation
   measured at the critical corrugation $W_{0c}\simeq0.44$.
 }
 \label{fig:ThetaLoc}
 \end{center}
\end{figure}
The drastic energy changes taking place across the transition have a
clear structural origin.
As was the case at $T=0$,
the moir\'e pattern conserves
its shape and symmetry across the unpinned-pinned transition, but the
central domains enclosed by the honeycomb-shaped network of domain walls
undergo a sharp structural transformation.\cite{mandelli2015prb}
At pinning, the portion of $2D$ lattice inside each hexagon rotates 
transforming from misaligned and incommensurate to approximately
aligned and commensurate with the underlying periodic potential.
This transformation can be followed by calculating the average local
lattice orientation of the colloidal monolayer defined as
\begin{equation}
\label{eq:thetaloc}
\theta_{\rm loc}=\langle\frac 1M \sum_{\langle i,j \rangle}
\bmod\!\left(\theta_{ij},\frac \pi3\right)\rangle
,
\end{equation}
where the sum is over all $M$ pairs $\langle i,j \rangle$ of nearest
neighbor particles with coordination six (excluding therefore the soliton regions), 
and $\theta_{ij}$ is the angle between the relative
position vector ${\bf r}_i-{\bf r}_j$ and the $x$-axis.
Figure \ref{fig:ThetaLoc} reports $\theta_{loc}$ as a function of
temperature for both the pinned and unpinned phases.
It is clear that below $T=T_c\simeq0.11$, and as the corrugation increases
across the transition, all local hexagonal domains between solitons locally rotate away from the initial
Novaco-Mc Tague orientation and back in approximate registry with the substrate.
The sharp drop of particle-potential energy $W$ that was seen to take place
at pinning corresponds precisely to the falling of most particles inside
each hexagonal cell into potential minima,
an event
which occurs at local commensurability. 
\section{Unpinned-Pinned Critical Point}
\label{sec:CP}
\begin{figure}[!t]
 \begin{center}
 \includegraphics[angle=0, width=0.45\textwidth]{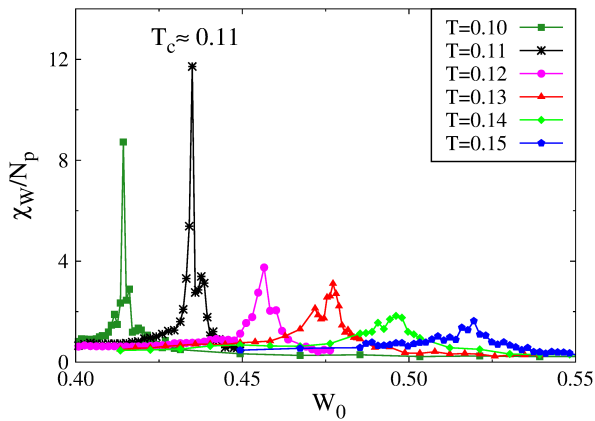}
 \caption{(Color online)
   The susceptibility $\chi_W$ defined in Eq.~\eqref{eq:ChiW}
   as a function of the substrate corrugation for different temperatures.
   A sharp peak appears when approaching $T_c\simeq0.11$, confirming the
   presence of a critical point.
 }
 \label{fig:ChiW}
 \end{center}
\end{figure}
Results of Figs.\ref{fig:deltaNsaddle} and \ref{fig:ThetaLoc} show that
first-order discontinuities connected with the unpinned-pinned transition
diminish with increasing temperature, until they vanish near
$T=T_c\simeq0.11$, at $W_0=W_{0c}\simeq0.44$.
These parameters appear to identify a novel $2D$ critical point.
In order to ascertain criticality we study the susceptibility
\begin{eqnarray}
  \chi_W &=&\frac{\partial \langle \sum_i w({\bf r}_i) \rangle}{\partial W_0}\nonumber\\
         &=&\frac{\langle \left(\sum_i w({\bf r}_i)\right)^2\rangle - \langle \sum_i w({\bf r}_i)\rangle^2}{k_BT},
\label{eq:ChiW}
\end{eqnarray}
which is obtained from the thermal fluctuations of the (dimensionless) substrate potential energy
$W/W_0=\langle \sum_i w({\bf r}_i)\rangle$.

\begin{figure}[!t]
 \begin{center}
 \includegraphics[angle=0, width=0.45\textwidth]{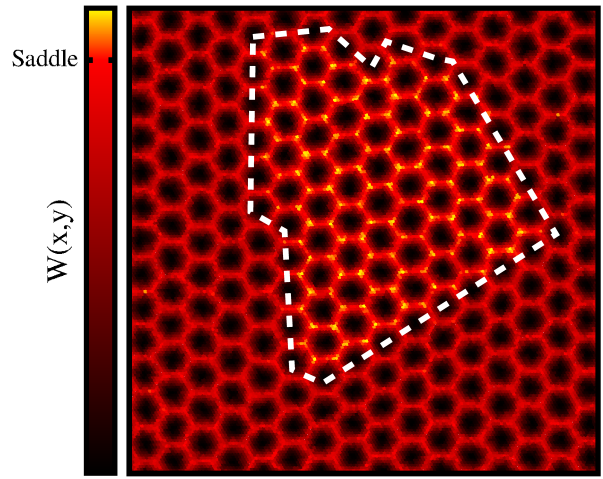}
 \caption{(Color online)
   Snapshot of a simulation performed at $T=0.04$ and $W_1<W_0=0.294<W_2$,
   within the coexistence region.
   Each particle is colored according to the value of the underlying
   corrugation potential $W(x,y)$: dark for positions close to the minima,
   and light colors for positions close to the maxima of $W(x,y)$.
   The dashed line highlights a large portion of the simulation supercell
   where the colloidal lattice is in the unpinned phase.
   This can be seen by the presence of particles (bright yellow dots)
   residing in energetically unfavourable positions very close to the
   maxima of the substrate potential.
   The rest of the system is instead in the pinned phase, characterized by
   a nearly complete absence of particles residing close to the maxima.
 }
 \label{fig:coexistence}
 \end{center}
\end{figure}

The result in Fig.~\ref{fig:ChiW} shows a sharp susceptibility peak at
$T_c$, confirming the presence of a critical point.
A finite-temperature critical Aubry transition is a major novelty predicted
for this system. The nature of this critical point is quite interesting,
and can be rationalized by analogy with the gas-liquid critical point.
Just below the gas-liquid critical point, droplets of liquid coexist with
large gas bubbles in the two-phase region.
As the critical point is approached, the boundaries between gas and liquid
get fuzzier and increasingly fluctuating, with length scales which
eventually diverge.
Here the situation is similar. 
Portions of the moir\'e honeycomb remain unpinned and even locally
misaligned, others turn toward zero local angle and become pinned, as
Fig.~\ref{fig:coexistence} shows.
Eventually, their fluctuating and fuzzy boundaries of increasing size make
the separation less and less clear until it disappears at the critical
point.

What critical indices should this new critical point have? We try to
anticipate the outcome by means of universality, which is based on
symmetry.
In a misaligned monolayer, the unpinned and the pinned states appear to share the same space group
symmetry.
In addition, once the global misalignment angle is fixed, thus taking care
of all $60^\circ$ rotations, there is no further symmetry left in either
phase.
This makes the analogy with gas-liquid quite strong, suggesting that the
unpinned-pinned critical point should be Ising-like.
Present size and time limitations do not permit the extraction of critical
indices from our simulations, and that task will remain for further work.
\section{Static Friction}
The two monolayer phases below $T_c$ are the finite temperature continuations of the unpinned 
and pinned phases already studied at $T=0$. As in that case, they are expected to exhibit respectively 
zero and finite static friction, defined as the minimal applied force that can cause sliding. 

\begin{figure}[!t]
 \begin{center}
 \includegraphics[angle=0, width=0.45\textwidth]{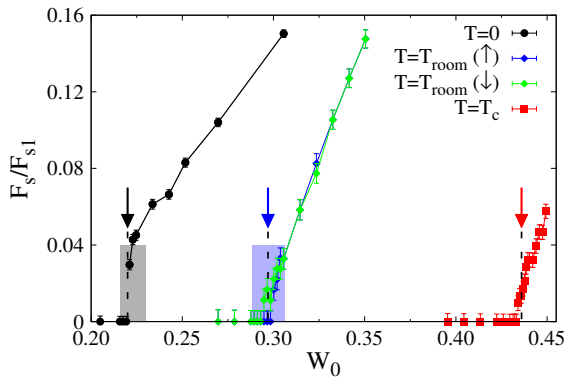}
 \caption{(Color online)
   The static friction force $F_{s}$, normalized with respect to
   the single particle value $F_{s1}=8\pi W_0/9a_l$ as a function of $W_0$ for three different
   temperatures $T$ = 0, 0.04 ($\sim T_{room}$), 0.11 ($\sim T_c$).
Shaded areas indicate the two-phase coexistence region extracted from the phase diagram of Fig.\ref{fig:phase-diagram}.
The two curves at $T=T_{room}$ are obtained upon increasing (decreasing)  $W_0$, and are
not meaningful inside the two-phase region, where the monolayer is inhomogeneous.
   The black, blue and red arrows indicate the potential magnitudes $W_0^*\simeq0.22$, 0.30, and $W_{0c}\simeq0.44$ 
respectively of the first order transitions at $T=0$ and $T=0.04$, and of the $T=0.11$ critical point. 
Note that the static friction jump, visible at $T=0$, is still present at  $T$ = 0.04 (even if artificially smeared out by two-phase coexistence),
while it disappears at  $T_c \sim 0.11$, a temperature where static friction becomes continuous versus $W_0$.  
Error bars represent the  $\Delta F$ step adopted in the protocol with increasing external homogeneous force $F_d$.
 }
 \label{fig:depinning}
 \end{center}
\end{figure}

Fig.\ref{fig:depinning} shows the static friction results at $T$ = 0, 0.04 ($\sim T_{room}$), and 0.11 ($\sim T_c$),
obtained as explained in Sec.\ref{sec:Mod&Prot}.
The lubric nature of the small $W_0$ phase and the more frictional nature of the large $W_0$ phase are 
confirmed. However the large static friction jump at $T=0$ between the two phases is generally smeared with 
temperature, until at $T\sim T_c$ static friction appears already somewhat below the transition.

It can be expected that this overall behavior of phases, probably with a sharp transition from viscous 
friction to stick-slip should carry over to dynamic friction. This aspect will form the object of a future study. 
\section{Discussion and Conclusions}
\label{sec:Concl}
The model study presented in this work establishes that a sharp
unpinned-pinned transition for increasing periodic potential acting on an
incommensurate lattice of particles, first established by Aubry in $1D$
where it is of second order and strictly at $T=0$, should carry over in
$2D$ at $T>0$.
For the particular case of an incommensurate colloid system which we model
here, the transition can be even sharper, first order instead of second
order, and extending to realistically finite temperatures.
The transition is structural, with portions of the moir\'e turning from
locally misaligned to aligned.
The pinned phase correspondingly gains energy at the transition, while at
the same time both interparticle energy and entropy suffer a
corresponding loss, as shown by the large slope of the coexistence phase
line.
The phase line ends in a critical point, where the unpinned-pinned
distinction disappears and fluctuations appear to diverge.

While obtained for a specific model and incommensurability, these results
should qualitatively persist for more general parameter values.
The magnitude of first-order jumps is connected with that of the Novaco
angle, in turn related to the value of incommensurability $\rho$.
By choosing $\rho<0.927$, the first-order character, and with that the
width of the two-phase coexistence region and the value of $T_c$, will
increase.
The frictional behavior changes from lubricity to pinning at the transition.
The dynamical friction in the coexistence region constitutes an
interesting question for further work.

The novel predicted critical point can and should be accessible
experimentally.
In fact, different incommensurabilities will imply different critical temperatures.
Therefore, even if experimental temperature is by necessity fixed at its room value,
a choice of $\rho$ closer and closer to one can always be found, where $T_c\sim T_{room}$,
making the critical point fully accessible.

It will also be interesting in the future to study the nature and
properties of the unpinned-pinned Aubry transition in, e.g., $2D$ systems
different from colloid monolayers, such as could be realized by compressing
two sheet materials together, or by modifying the adhesive interaction of
$2D$ adsorbates layers by charging.
The nature of these systems is sufficiently different from colloid
monolayers to suggest that there might be substantial differences, as well as analogies.
In cases where the Novaco-McTague misalignment does not occur, all
first-order characters of the transition might be weaker; but its existence
at finite temperature should at least persist.
We also expect that the precise nature of interparticle interactions will make a quantitative, but probably not a total difference.  The 2D Aubry transition should persist for example in systems where interparticle interactions have an attractive part, so long as these do not lead to a  2D lattice collapse. The 2D Frenkel-Kontorova model in particular,\cite{braun_kivshar} still to be studied in this respect, should show an Aubry-type transition as well. 

\acknowledgments
We are grateful to T. Brazda and C. Bechinger for much ongoing exchange of
ideas and information, and to A. Silva and R. Guerra for discussion.
This work was mainly supported under the ERC Advanced Grant
No.\ 320796-MODPHYSFRICT, and by COST Action MP1303.



\begin{thebibliography}{99}
%
\bibitem{vanossiRevModPhys}
 A. Vanossi, N. Manini, M. Urbakh, S. Zapperi, and E. Tosatti,  Rev. Mod. Phys. {\bf 85}, 529 (2013).
%
\bibitem{Manini16} 
  N. Manini, O.M. Braun, E. Tosatti, R. Guerra, and A. Vanossi,
  J. Phys.: Condens. Matter {\bf 28}, 293001 (2016).
%
\bibitem{dienwiebel2004}
  M. Dienwiebel, G. S. Verhoeven, N. Pradeep, J. W. M. Frenken,
  J.A. Heimberg, and H.W. Zandbergen, Phys. Rev. Lett. {\bf 92}, 126101
  (2004).
%
\bibitem{leven2013}
  I. Leven, D., Krepel, O. Shemesh, O. Hod, J. Phys. Chem. Lett. {\bf 4}, 115-120 (2013).
%
\bibitem{pierno2015}
  M. Pierno, L. Bruschi, G. Mistura, G. Paolicelli, A. di Bona,
  S. Valeri, R. Guerra, A. Vanossi, and E. Tosatti, Nat. Nanotech.
  {\bf 106}, 1 (2015).
%
\bibitem{bohlein2012}
  T. Bohlein, J. Mikhael, and C. Bechinger, Nat. Mater. {\bf 11}, 126 (2012).
%
\bibitem{vanossi2012}
  A. Vanossi, N. Manini, and E. Tosatti, Proc. Natl. Acad. Sci. USA {\bf 109}, 16429 (2012).
%
\bibitem{Floria96}
  L.M. Flor\'\i{}a and J.J. Mazo, Adv. Phys. {\bf 45}, 505 (1996).
%
\bibitem{braun_kivshar}
  O. M. Braun and Y. Kivshar, {\it The Frenkel-Kontorova Model:
    Concepts, Methods, and Applications} (Springer, Berlin, 1998).
%
\bibitem{aubry1983}
  S. Aubry and P. Y. Le Daeron, Physica D {\bf 8}, 381 (1983).
%
\bibitem{coppersmith1983}
  S. N. Coppersmith and D. S. Fisher, Phys. Rev. B {\bf 28}, 2566 (1983).
%
\bibitem{mazzucchelli1985}
  G. M. Mazzucchelli and R. Zeyher, Z. Phys. B {\bf 62}, 367 (1985).
%
\bibitem{benassi2011}
 A. Benassi, A. Vanossi, E. Tosatti, Nat. Commun. {\bf 2}, 236 (2011).
%
\bibitem{mandelli2013}
 D. Mandelli, A. Vanossi, E. Tosatti, Phys. Rev. B {\bf 87}, 195418 (2013).
%
\bibitem{gangloff2015}
 D. Gangloff, A. Bylinskii, I. Counts, W. Jhe, V. Vuleti\'c, Nat. Phys. {\bf 11}, 915 (2015).
%
\bibitem{bylinskii2015}
 A. Bylinskii, D. Gangloff, V. Vuleti\'c, Science {\bf 348}, 1115 (2015). 
%
\bibitem{mandelli2015prb}
  D. Mandelli, A. Vanossi, M. Invernizzi, S. Paronuzzi, N. Manini and E. Tosatti, Phys. Rev. B {\bf 92}, 134306 (2015).
%
\bibitem{Reichhardt16} 
C. Reichhardt and C.J.O. Reichhardt, Rep. Prog. Phys. {\bf 80}, 026501 (2016).
%
\bibitem{novaco1977}
  A. D. Novaco and J. P. Mc Tague, Phys. Rev. Lett. {\bf 38}, 1286 (1977).
%
\bibitem{trambly2010}
  G. Trambly de Laissardi\`ere, D. Mayou, and L. Magaud, Nano Lett.
  {\bf 10}, 804 (2010).
%
\bibitem{mandelli2015prl}
  D. Mandelli, A. Vanossi, N. Manini, and E. Tosatti, Phys. Rev. Lett. {\bf 114}, 108302 (2015).
%
\bibitem{bohr1992}
  F. Grey and J. Bohr, Europhys. Lett. {\bf 18}, 717 (1992).
%
\bibitem{johnston2014}
 D. C. Johnston, arXiv:1402.1205v1 [cond-mat.soft].
%
\bibitem{henderson1969}
 M. Henderson, M. S. Wertheim, J. Chem. Phys. {\bf 51}, 5420 (1969).
%
\end{thebibliography}
\end{document}